\documentclass[amsmath,aps,twocolumn]{revtex4}

\usepackage{graphicx}
\usepackage{bm}

\renewcommand{\Re}{\operatorname{Re}}

\begin{document}
\title{Identifying reactive trajectories using a moving transition state}
\author{Thomas Bartsch}
\author{T. Uzer}
\affiliation{Center for Nonlinear Science,
  Georgia Institute of Technology, Atlanta, GA 30332-0430, USA}
\author{Jeremy M. Moix}
\author{Rigoberto Hernandez}
\affiliation{Center for Computational Molecular Sciences \& Technology,
  School of Chemistry \& Biochemistry,
  Georgia Institute of Technology, Atlanta, GA 30332-0430, USA}
\date{\today}

\begin{abstract}

A time-dependent no-recrossing dividing surface is shown to lead to a new
criterion for identifying reactive trajectories well before they are
evolved to infinite time.  Numerical dynamics simulations of a dissipative
anharmonic two-dimensional system confirm the efficiency of this approach.
The results are compared to the standard fixed transition state dividing
surface that is well-known to suffer from recrossings and therefore
requires trajectories to be evolved over a long time interval before they
can reliably be classified as reactive or non-reactive.  The moving
dividing surface can be used to identify reactive trajectories in harmonic
or moderately anharmonic systems with considerably lower numerical effort
or even without any simulation at all.

\end{abstract}
\maketitle

\section{Introduction}

It is perhaps surprising how many problems in chemistry, physics, and
biology can be reduced to the simple model of diffusion over a
barrier~\cite{Talkner95a}.
Although chemical reactions in all phases of matter provide the
prime example \cite{Miller,truh96}, a plethora of systems that evolve from
suitably defined ``reactant'' to ``product'' states are amenable to a
description in this framework
\cite{genesis,Koon00,Komatsuzaki02,JFU00,ballistic,toller}.  Transition
State Theory (TST) \cite{truh85,rmp90,truh96} or one of its descendants
\cite{grot80,pgh89,mm86} is often used to approximate the rate of these
reaction processes.  
This theory is based on the assumption that the reaction rate
is determined in a small volume of the phase space near the barrier.  It is
then possible to define a dividing surface separating reactants from
products and obtain the rate from the flux through this surface.  The
optimal location of the dividing surface is that which minimizes the number
of recrossings---the fundamental idea of variational transition state
theory \cite{truh84,pollak90a,frish92,tucker92,tucker95}.
When the system of interest can be viewed as isolated from
its environment, as in low-density gas phase chemical reactions, TST may
indeed provide an excellent approximation to the rate.  However, most
processes of interest do not take place in isolation, but rather in a
complex environment where interactions between the system and its
surroundings occur on time scales comparable to that of the reaction: In a
reaction occurring in the condensed phase, for instance, the dynamics of
the solute is strongly coupled to that of the solvent.  In this case, the
fundamental assumption of TST, that the dividing surface is crossed once
and only once, no longer holds \cite{sbb85,sbb86,sbb88,talkner95}.  On the
time scale of the reaction event, fluctuations of the environment will
almost inevitably cause recrossings of the dividing surface that lead to an
overestimation of the rate.

To overcome the recrossing problem, many TS dividing surfaces
have been suggested in the literature
\cite{hynes85a,truh96,pollak86b,tucker95}
which provide systematic (and simple) approximations to 
the optimal TS dividing surface.
In some cases, the dividing
surface has been identified in the infinite-dimensional phase space 
consisting of the system and an explicit
set of bath oscillators \cite{zwan73,caldeira81,pollak86b}. 
This approach leads to an excellent
approximation to the rate \cite{pollak86b,pgh89}. 
Interestingly, the same result was subsequently obtained
without recourse to the explicit heat bath model,
using instead a collective reaction coordinate 
containing the influence of the bath directly \cite{Graham90}.

In a recent series of papers \cite{dawn05,halcyon}, we reformulated the
recrossing problem using a dividing surface that is itself
moving stochastically so as to avoid recrossings. 
The motion of that surface follows the unique trajectory
---named the TS trajectory--- that never leaves the barrier region. 
Any reactive trajectory crosses the moving surface once and only once, 
whereas a nonreactive trajectory does not cross at all. This construction
extends the approach of \cite{Graham90} in that it provides not only a
reaction coordinate, but also the complete geometric structure by
specifying all of the unstable and
stable degrees of freedom globally \cite{Martens02}.
The previous purely analytic studies \cite{dawn05,halcyon} are
complemented here with a numerical investigation of the reaction dynamics
for a two-dimensional stochastic nonlinear model.  It will be shown that the
moving dividing surface offers considerable computational advantages over
the traditional fixed surface: Its use can significantly reduce the
simulation time required to distinguish between reactive and nonreactive
trajectories.  Indeed, for a harmonic barrier it identifies reactive
trajectories \emph{a priori}, so that the need to simulate their dynamics
does not arise at all. In an anharmonic system, the identification of
reactive trajectories by the moving surface is no longer
exact. Nevertheless, for moderately strong anharmonicities it provides a
useful approximation, and its advantages over the fixed surface are
retained.  In addition, the moving TS surface introduces novel observables
that characterize the reaction process on a microscopic level. Most
prominently, it allows one to define a unique reaction time for each
individual trajectory.

The outline of this paper is as follows:
In Sec.~\ref{sec:T-M}, the two-dimensional stochastic nonlinear
model that is the focus of the computational studies in this
work is defined and the construction of 
the moving TS dividing surface and its associated geometric structures 
is briefly reviewed.
In Sec.~\ref{sec:prob}, the ensemble of trajectories 
is specified by a thermal distribution of particles
localized at the conventional TST dividing surface.
This barrier ensemble is reminiscent of the weighting
distribution in standard rate expressions
and is appropriate even in nonlinear cases.
Its simple structure also readily leads to the analytic determination of
several observables of the model system in the harmonic limit 
(Sec.~\ref{sec:anal}).
They are in precise agreement with the numerical results
presented in Sec.~\ref{sec:results}.
The latter section also demonstrates that observables converge
faster when evaluated using the moving dividing surface rather 
than conventional numerical methods, 
both in the harmonic limit and in systems with anharmonic barriers.
This observation is particularly useful in the anharmonic case when the
chosen system is not amenable to analytic approaches.

\section{Preliminaries}\label{sec:T-M}

Although the general theory is applicable to systems with
an arbitrary number $n$ of 
degrees of freedom, the following discussion
will be restricted to $n=2$ coordinates under the influence of a stochastic bath. 
This choice can be made without loss of
generality because it exhibits all the salient features of the
higher-dimensional cases:
It can encompass an unbound (reactive) direction and a bound bath mode whose
interaction with the reactive mode is strong enough to require its
explicit description.
The coupling of the modes is described by
a Taylor expansion about a transition point (or col) on
the potential energy surface.
Such a model with a minimum number of nonlinear terms is
described in this section. It will be used in the following
to study the effect of
increasing anharmonicity on the identification of reactive trajectories.

To set the stage for the following investigations, 
the construction of the moving TST dividing surface and the
associated invariant manifolds is summarized in the remainder of this 
section.  The reader interested in a full
exposition is referred to Refs.~\onlinecite{dawn05}
and \onlinecite{halcyon}, 
where the formalism was
first introduced.

\subsection{The Two-Dimensional Dissipative Model}

A prototypical reactive system within a solvent may be
described by the Langevin equation \cite{zwan01}
\begin{equation}
  \label{LE}
  \ddot{\vec q}_\alpha(t) = -\nabla_{\vec q} U(\vec q_\alpha(t)) -
        {\bm\Gamma}\dot{\vec q}_\alpha(t) +
        \vec\xi_\alpha(t) \;.
\end{equation}
The vector $\vec q$ here denotes a set of $n=2$ mass-weighted coordinates,
$U(\vec q)$ the potential of mean force governing the reaction, $\bm\Gamma$
a symmetric positive-definite friction matrix and $\vec\xi_\alpha(t)$ a
fluctuating force assumed to be Gaussian with zero mean.  The subscript
$\alpha$ represents randomness by labeling different instances of the
fluctuating force. The latter is related to the friction matrix $\bm\Gamma$
by the fluctuation-dissipation theorem \cite{zwan01}
\begin{equation}
  \label{FDTWhite}
  \left<\vec\xi_\alpha(t)\vec\xi_\alpha^\text{T}(t')\right>_\alpha = 
      2\,k_\text{B}T\, {\bm\Gamma} \, \delta(t-t') \;,
\end{equation}
where the angular brackets denote the average over the instances $\alpha$
of the noise.
Although not strictly necessary, 
the friction is often taken to be isotropic, {\it i.e.},
\begin{equation}
  \label{isoFrict}
  \bm{\Gamma}=\gamma\bm{I} \;,
\end{equation}
with a scalar friction constant $\gamma$.

The reactant and product regions in configuration space are separated by a
potential barrier whose position is marked by a saddle point $\vec
q_0^\ddag=0$ of the potential $U(\vec q)$.  In its vicinity, the potential
is approximately harmonic and can always be written in a diagonal normal
form. In general, anharmonic terms will be present in the
potential. In the neighborhood of the
saddle point, where the reaction rate is determined, they are only
moderately strong, but  usually
not negligible. In this work, we include a typical (even) higher-order
nonlinear term and focus on the potential
\begin{equation}
U(x,y)=-\frac{1}{2}\omega_x^2 x^2+\frac{1}{2}\omega_y^2 y^2 
+ k x^2y^2
\;,
\label{eq:pot}
\end{equation}
where the position vector is written as $\vec q=(x,y)$, and the constant $k$ quantifies
the nonlinear coupling of the different degrees of freedom.
Note that the nonlinearity in the potential (\ref{eq:pot}) is symmetric in
the coordinate system and neglects other fourth order terms that are typically
retained in the analysis of anharmonic barriers.  (See, e.g.,
Ref.~\onlinecite{hern94}, in which such coupled anharmonic
potentials have been used to study the 
${\rm H} + {\rm H}_2 \rightleftharpoons {\rm H}_2 + {\rm H}$
reaction and bound vibrational systems.)
However, as discussed in the Appendix, it is amenable to an analytic treatment
that simplifies the numerical computation of the forward and backward 
trajectories, while providing sufficient coupling to break the 
exact integrability of the harmonic system.

In the special case $k=0$, the system is globally harmonic. In this case,
the constructions outlined below yield a moving dividing surface that is
strictly free of recrossings. If $k\ne 0$, deviations from the harmonic dynamics
will arise outside the TS region that may lead to error in the identification of
reactive trajectories.  Nonetheless, the wealth  of microscopic detail that
the moving dividing surface reveals can most easily be
illustrated using the harmonic limit. This is shown in Secs.~\ref{sec:prob}
and~\ref{ssec:resultsHarm}.  The real power of the numerical method,
however, lies in addressing nonlinear systems; the accuracy of the
approximate identification of nonlinear reactive trajectories is discussed
in Sec.~\ref{ssec:resultsNonlin}.

With the potential~(\ref{eq:pot}), the Langevin equation~(\ref{LE}) reads
\begin{equation}
  \label{linLE}
  \ddot{\vec q}_\alpha(t) = {\bm\Omega} \vec q_\alpha(t)
     + O(q_\alpha^3)
     - {\bm\Gamma}\dot{\vec q}_\alpha(t) + \vec\xi_\alpha(t) \;,
\end{equation}
where
\begin{equation}
  \bm{\Omega} = \begin{pmatrix}
                  \omega_x^2 & 0 \\
                  0 & -\omega_y^2
                \end{pmatrix}
\end{equation}
is the matrix of second derivatives of $U(\vec q)$. 
The nonlinear terms in~(\ref{linLE}), which stem from the anharmonic
contributions to the potential~(\ref{eq:pot}) will be ignored in the
remainder of this
section, where an exact dividing surface for the harmonic limit will be
constructed. The full nonlinear equation of motion~(\ref{LE}) will be taken
up again in the numerical calculations of Sec.~\ref{ssec:resultsNonlin}.
The following presentation can easily be generalized to $N$ spatial
dimensions if $y$ is understood to denote
an ($n-1$)-dimensional vector and the corresponding squared frequency
$\omega_y^2$ an ($n-1$)-dimensional symmetric matrix whose
eigenvalues need not be degenerate.

\subsection{The Transition State Trajectory}

As was shown in~\cite{dawn05,halcyon}, 
Eq.~(\ref{linLE}) can be rewritten in phase space,
$\vec z=(\vec q, \vec v)$, with $\vec v=\dot{\vec q}$, as
\begin{equation}
  \label{phaseLE}
  \dot{\vec z}_\alpha(t)=A \vec z_\alpha(t)+
                        \begin{pmatrix}
                           0 \\
                           \vec\xi_\alpha(t)
                        \end{pmatrix}
\end{equation}
with the $2n$-dimensional constant matrix
\begin{equation}
  \bm{A}=\begin{pmatrix}
          \bm{0} & \bm{I}\\
          \bm{\Omega} & -\bm{\Gamma}
    \end{pmatrix}\;,
\end{equation}
where $\bm{I}$ is the $n\times n$ identity matrix.  The matrix $\bm{A}$ is
readily diagonalized to yield the eigenvalues $\epsilon_j$ and the
corresponding eigenvectors $\vec V_j$.  Equation~(\ref{phaseLE}) then
decomposes into a set of $2n$ independent scalar equations of motion
\begin{equation}
  \label{diagEq}
  \dot z_{\alpha j}(t) = \epsilon_j z_{\alpha j}(t) + \xi_{\alpha j}(t) \;,
\end{equation}
where $z_{\alpha j}$ are the components of $\vec z$ in the basis $\vec V_j$
of eigenvectors of $\bm{A}$ and $\xi_{\alpha j}$ are the corresponding
components of $(0,\vec\xi_\alpha(t))$.

A particular solution of Eq.~(\ref{diagEq}) is given by
\begin{equation}
z_{\alpha j}^{\ddag}(t)=
    \begin{cases}
 \phantom{-}\int_{-\infty}^0 e^{-\epsilon_j \tau}
                        \xi_{\alpha j}(t+\tau)\, d\tau\\
            \hspace{6em} \text{if $j$ such that $\Re\epsilon_j<0$},\\[2ex]
         -\int_0^{\infty} e^{-\epsilon_j \tau}
                         \xi_{\alpha j}(t+\tau)\, d\tau\\
            \hspace{6em} \text{if  $j$ such that $\Re\epsilon_j>0$.}
    \end{cases}
\label{eq:TSTtraj}
\end{equation}
Whereas a typical trajectory will eventually descend into either the
reactant or the product wells, the trajectory given by Eq.~(\ref{eq:TSTtraj})
has the important property \cite{dawn05,halcyon} that it
remains in the vicinity of the saddle point for all time. In this respect
it resembles the equilibrium position on the saddle that represents the
unique trajectory in the absence of noise that never descends from the
saddle. We named this distinguished trajectory the Transition State
Trajectory in \cite{dawn05,halcyon} because it plays as central a role in the TST
in a noisy environment as the equilibrium point does in conventional
TST. Although the integral
representation~(\ref{eq:TSTtraj}) defines the TS trajectory, it does not
provide the most efficient way of calculating it. In fact, by means of an
algorithm that we introduced in \cite{halcyon} an
instance of the TS trajectory can be sampled almost as efficiently as an
instance of the fluctuating force itself.

\subsection{The relative dynamics}

Once the TS Trajectory $\vec z_\alpha^\ddag(t) = (\vec q_\alpha^\ddag(t),
\vec v_\alpha^\ddag(t))$ is given, any other trajectory under the influence
of the same noise can be described in relative coordinates
\begin{equation}
  \label{relCoord}
  \Delta\vec z(t) = \begin{pmatrix}
                       \Delta\vec q(t) \\ \Delta\vec v(t)
                    \end{pmatrix}
                  = \vec z_\alpha(t) - \vec z_\alpha^\ddag(t) \;,
\end{equation}
where the TS Trajectory serves as the origin of a moving coordinate
system. The relative coordinate vectors can be written
without a subscript $\alpha$ because they satisfy the noiseless equation of
motion
\begin{equation}
  \Delta\ddot{\vec q}(t) = {\bm\Omega}\, \Delta\vec q(t) -
        {\bm\Gamma}\,\Delta\dot{\vec q}(t) \;,
\end{equation}
or, in phase space,
\begin{equation}
  \Delta\dot{\vec z}(t) = \bm{A}\,\Delta\vec z(t)
\end{equation}
and are, therefore, independent of the noise.  Using the eigenvectors of
$\bm{A}$, one can construct invariant manifolds and a no-recrossing surface
of the noiseless relative dynamics. According to Eq.~(\ref{relCoord}), they can
then be regarded as being attached to the TS Trajectory and being carried
around by it. In this way one obtains moving invariant manifolds and a
moving no-recrossing surface in the phase space of the full, noisy dynamics
\cite{dawn05,halcyon}.

In the two-dimensional case of the potential in Eq.~(\ref{eq:pot})
under isotropic friction as specified in Eq.~(\ref{isoFrict}),
the eigenvalues of $\bm{A}$ can be found explicitly:
\begin{alignat}{2}
\epsilon_\text{u}  &=  
          -\frac 12\left(\gamma-\sqrt{\gamma^2+4\omega_x^2}\right)\;,\nonumber\\
\epsilon_\text{s}  &=  
          -\frac 12\left(\gamma+\sqrt{\gamma^2+4\omega_x^2}\right)\;,\nonumber\\
\epsilon_\text{t1}  &=  
          -\frac 12\left(\gamma-\sqrt{\gamma^2-4\omega_y^2}\right)\;,\nonumber\\ 
\epsilon_\text{t2}  &=  
          -\frac 12\left(\gamma+\sqrt{\gamma^2-4\omega_y^2}\right)\;.
\end{alignat}
The corresponding eigenvectors read
\begin{alignat}{4}
\label{evecs}
\vec V_\text{u} &= \begin{pmatrix}
                     1 \\ 0 \\ \epsilon_\text{u} \\ 0
                   \end{pmatrix} \;,\qquad
\vec V_\text{s} &= \begin{pmatrix}
                     1 \\ 0 \\ \epsilon_\text{s} \\ 0
                   \end{pmatrix} \;,\nonumber\\
\vec V_\text{t1} &= \begin{pmatrix}
                     0 \\ 1 \\ 0 \\ \epsilon_\text{t1}
                   \end{pmatrix} \;,\qquad
\vec V_\text{t2} &= \begin{pmatrix}
                     0 \\ 1 \\ 0 \\ \epsilon_\text{t2}
                   \end{pmatrix} \;.
\end{alignat}
These simple analytic results are obtained because iso\-tro\-pic friction leads
to a decoupling of the reactive and the transverse degrees of freedom. The
eigenvectors $\vec V_\text{u}$ and $\vec V_\text{s}$ span the reactive
$x$-$v_x$ subspace, whereas $\vec V_\text{t1}$ and $\vec V_\text{t2}$ span
the transverse subspace $y$-$v_y$.

The knowledge of the eigenvectors allows one to explicitly specify the
coordinate transformation between position-velocity coordinates $\Delta x,
\Delta y, \Delta v_x, \Delta v_y$ and the diagonal coordinates $\Delta
x_\text{u}, \Delta x_\text{s}, \Delta x_\text{t1}, \Delta x_\text{t2}$ that
characterize a phase space point via $\Delta \vec x = \sum_i \Delta x_i
\vec V_i$. In the reactive subspace, these transformations read
\begin{equation}
  \Delta x=\Delta x_\text{u} + \Delta x_\text{s} \;, \qquad
  \Delta v_x = \epsilon_\text{u}\,\Delta x_\text{u} 
             + \epsilon_\text{s}\,\Delta x_\text{s}
\end{equation}
with the inverse
\begin{equation}
  \Delta x_\text{u} = \frac{\Delta v_x-\epsilon_\text{s}\,\Delta x}
                           {\epsilon_\text{u}-\epsilon_\text{s}} \;,\qquad
  \Delta x_\text{s} = \frac{-\Delta v_x+\epsilon_\text{u}\,\Delta x}
                           {\epsilon_\text{u}-\epsilon_\text{s}} \;.
\end{equation}

\begin{figure}
 \centerline{\includegraphics[width=\columnwidth]{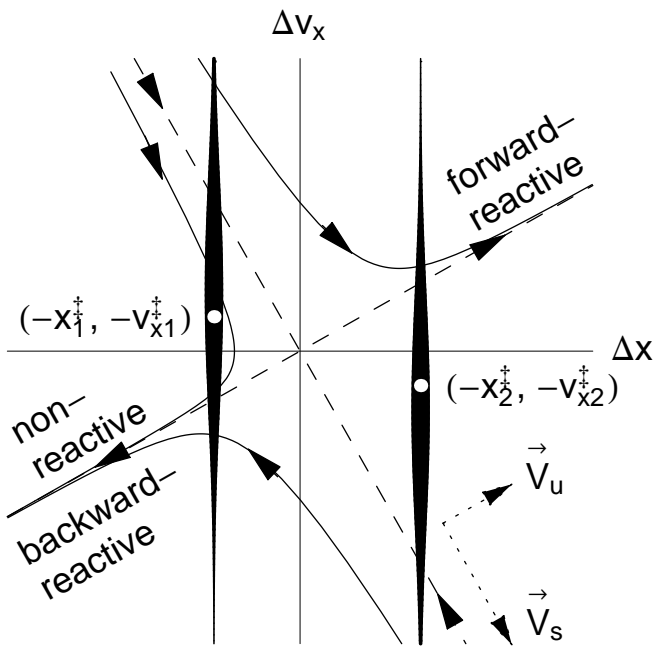}}
\caption{\label{fig:ensemble}
  Phase portrait illustrating the relative dynamics in the reactive degree
  of freedom. Dashed lines indicate the stable and unstable manifolds of
  the equilibrium point, and the dotted arrows display 
  the corresponding eigenvectors that
  span the diagonal coordinate system. Solid curves illustrate representative
  trajectories.
  White dots indicate two possibilities for the position and velocity of
  the TS trajectory at
  time $t=0$, the vertical lines the corresponding barrier
  ensembles. The probability density is given by the line widths.
}
\end{figure}

For all values of $\gamma$ and $\omega_x$, the eigenvalue
$\epsilon_\text{u}$ is positive, whereas $\epsilon_\text{s}$ is
negative. They correspond to one-dimensional stable and unstable subspaces
within the reactive degree of freedom which, together with representative
trajectories, are illustrated in Fig.~\ref{fig:ensemble}. The coordinate
$\Delta x_\text{u}$ determines the fate of a trajectory in the remote
future: Trajectories with $\Delta x_\text{u}>0$ descend into the product
well, those with $\Delta x_\text{u}<0$ into the reactant well. Similarly, a
stable coordinate $\Delta x_\text{s}>0$ indicates a trajectory that comes
out of the product well in the distant past, whereas a trajectory with
$\Delta x_\text{s}<0$ comes out of the reactant well. A forward-reactive
trajectory that changes from reactants to products is thus characterized by
$\Delta x_\text{s}<0$ and $\Delta x_\text{u}>0$, whereas a
backward-reactive trajectory has $\Delta x_\text{s}>0$ and $\Delta
x_\text{u}<0$. Each reactive trajectory crosses the line $\Delta x=0$ once
and only once. This line, or in several degrees of freedom the hypersurface
defined by this condition, can therefore serve as a recrossing-free
dividing surface between reactants and products. Furthermore, the invariant
stable and unstable subspaces themselves act as separatrices between
reactive and nonreactive trajectories. Once the initial condition of a
trajectory is known relative to these separatrices, it can unambiguously be
classified as reactive or non-reactive.

\section{The Barrier Ensemble}\label{sec:prob}

The rate calculation of infrequent events ---such as that in
chemical reactions--- can be greatly simplified by sampling
trajectories in the transition state region rather than in the
reactant region \cite{Keck62,Keck67,chan78,tully81,and90}.
The transition path sampling technique, for example, 
focuses exclusively on reactive trajectories
and therefore
mitigates the difficulty of studying high-dimensional
systems \cite{chan98a,chan99a,chan02a,chan02b,andricioaei05}.
Nonetheless, the rate of infrequent events has long been known
to be described by a flux-flux correlation function 
relative to a fixed dividing surface \cite{yamamoto,zwan65,chan78}.
(But see Ref.~\onlinecite{andricioaei06} for a recent enhanced-sampling
strategy to smooth the potential and thereby speed up the calculations.)
One difficulty in computing the correlation function, however, is 
the need for the simulated trajectories to be evolved for very long
times simply to determine which trajectories are reactive.
We will show below that the use of the time-dependent TST dividing surface
may allow one to resolve that question in a more efficient way by 
identifying the nature of the trajectory ---{\it viz.}~reactive or not---
at significantly earlier evolution times.

In order to sample the reactive trajectories efficiently, it is useful
to use initial conditions in which all
the particles are placed on the fixed TST dividing surface ($x=0$)
at $t=0$.
That choice guarantees that the trajectories will cross the
surface at least once, but it does not prevent them from recrossing it.  
Consistent with the Boltzmann weighting in the
flux-flux-correlation function \cite{yamamoto,zwan65,chan78}, the
initial conditions are
distributed along the stable transverse direction $y$ and the velocities
according to the probability density function,
\begin{multline}
  \label{BarrDist}
  f(x,y,v_x,v_y) = (2\pi\,k_\text{B}T)^{3/2}\, \omega_y\,\times\\
    \exp\left\{-(\omega_y^2 y^2+v_x^2+v_y^2)/2\,k_\text{B}T\right\}\,
    \delta(x) \;.
\end{multline}
This choice defines the barrier ensemble.
The integration of the Boltzmann-weighted flux of these states gives the
TST estimate of the numerator of the rate expression.  If all these states
were reactive and never recrossed (returned to) the fixed TST dividing
surface, this estimate would be exact.  The questions to be resolved below concern the
deviation of the true dynamics from this TST estimate. These question will
be investigated for both harmonic and anharmonic barriers. In all cases,
the initial conditions will be sampled from the same barrier
ensemble~(\ref{BarrDist}).

A stochastic trajectory is determined not only by its initial condition,
but also by the specific instance of the fluctuating force that is acting
upon it. 
In a full-fledged rate calculation, for example, an average has to be taken
over both the initial conditions and the noise. 
The focus of this paper, however, is the information that can 
be obtained about the microscopic reaction dynamics
using the moving TS surface.
For simplicitly, a particular instance of the noise has therefore been
used to illustrate most of the results.
Nevertheless, the calculations presented here
were repeated for several such noise sequences
always leading to the same qualitative conclusions and
thereby confirming that the results shown here are indeed typical.
(These are not shown here for the sake of brevity.)
While averages over the noise will tend to wipe out much of this 
microscopic detail, it is instructive to confirm the convergence of
the identification of trajectories in calculating averages.
In what follows, the average of the forward and backward reaction
probability will be used to illustrate the convergence and degree of accuracy 
achievable using the moving TS surface to identify the
reactive trajectories.

\section{Analytic results}
\label{sec:anal}

Although anharmonicities have to be addessed in a typical chemical system,
it is helpful to begin with the harmonic limiting case 
because it is amenable to an analytic treatment. 
On the one hand, the harmonic limit illustrates the level of
microscopic detail in which the moving TST method allows one to describe
the reaction mechanism. On the other hand, the analytic results derived
here provide a benchmark against which the performance of the numerical
calculations of Sec.~\ref{sec:results} can be assessed.

\subsection{Reaction probabilities}
\label{ssec:rProb}

The fate of individual trajectories in the barrier
ensemble~(\ref{BarrDist}) can easily be determined if their initial
conditions are transformed into relative coordinates.  The projection onto
the reactive degree of freedom is illustrated in
Fig.~\ref{fig:ensemble}. Since in space-fixed coordinates the barrier
ensemble is centered around $\vec q=\vec v=0$, the distribution function in
relative coordinates is peaked at the stochastic position $\Delta\vec
q=-\vec q_\alpha^\ddag(0)$, $\Delta\vec v=-\vec
v_\alpha^\ddag(0)$. It reads explicitly
\begin{widetext}
\begin{multline}
  \label{RelDist}
  f_\text{rel}(\Delta x,\Delta y,\Delta v_x,\Delta v_y) =
    (2\pi\,k_\text{B}T)^{3/2}\, \omega_y\, \times\\
    \exp\left\{-(\omega_y^2 (\Delta y+y_\alpha^\ddag)^2
                 +(\Delta v_x+v_{x\alpha}^\ddag)^2
                 +(\Delta v_y+v_{y\alpha}^\ddag)^2)/2\,k_\text{B}T\right\}\,
    \delta(\Delta x+x_\alpha^\ddag) \;.
\end{multline}
\end{widetext}

The forward-reactive part of the ensemble is formed by those trajectories
whose initial velocity $\Delta v_x$ is so large that the trajectory lies
above both the stable and the unstable manifold of the equilibrium point.
The knowledge of the eigenvectors~(\ref{evecs}) allows one to locate these
separatrices quantitatively.  Reactive trajectories are thus found to be
characterized by the condition
\begin{equation}
  \Delta v_x > \Delta v_{x,\text{min}} :=
    \begin{cases}
       -x_\alpha^\ddag \epsilon_\text{s} & :\quad x_\alpha^\ddag>0\;, \\
       -x_\alpha^\ddag \epsilon_\text{u} & :\quad x_\alpha^\ddag<0\;.
    \end{cases}
\end{equation}
Therefore, the probability for a member of the barrier ensemble to be
forward-reactive is given by
\begin{align}
  \label{Preact}
  P_\text{f} &= 
    \int d\Delta x \int_{\Delta v_x>\Delta v_{x,\text{min}}} d\Delta v_x
   \,\times\nonumber\\
  &\hspace{5em}\int d\Delta y \int d\Delta v_y\;
        f_\text{rel}(\Delta x,\Delta y,\Delta v_x,\Delta v_y) \nonumber
\end{align}
or
\begin{align}
  P_\text{f} &= (2\pi\,k_\text{B}T)^{-1/2}\,\times\nonumber\\
  &\hspace{4em}
     \int_{\Delta v_{x,\text{min}}}^\infty d\Delta v_x
     \exp\left\{-(\Delta v_x+v_{x\alpha}^\ddag)^2/2\,k_\text{B}T\right\}
     \nonumber \\
  &= \frac 12 \operatorname{erfc}
     \left( \frac{\Delta v_{x,\text{min}}+v_{x\alpha}^\ddag}
                 {\sqrt{2\,k_\text{B}T}} \right) \;,
\end{align}
which has been written in terms of the complementary error function
\cite{Abramowitz}
\begin{equation}
  \operatorname{erfc}(x) = \frac{2}{\sqrt{\pi}}
    \int_x^\infty \exp(-t^2)\, dt \;.
\end{equation}
In a similar manner, backward-reactive trajectories satisfy
\begin{equation}
  \Delta v_x < \Delta v_{x,\text{max}} :=
    \begin{cases}
       -x_\alpha^\ddag \epsilon_\text{u} & :\quad x_\alpha^\ddag>0\;, \\
       -x_\alpha^\ddag \epsilon_\text{s} & :\quad x_\alpha^\ddag<0\;,
    \end{cases}
\end{equation}
and their probability in the ensemble is
\begin{equation}
  \label{Pback}
  P_\text{b}=
     \frac 12 \operatorname{erfc}
     \left(- \frac{\Delta v_{x,\text{max}}+v_{x\alpha}^\ddag}
                  {\sqrt{2\,k_\text{B}T}} \right) \;.
\end{equation}

\subsection{Reaction times}

In contradistinction to a space-fixed dividing surface, the moving 
TS surface is crossed once and only once by each reactive
trajectory. This allows us to define a unique reaction time $\Delta
t^\ddag$ for each reactive trajectory: It is the time when the trajectory
crosses the dividing surface, relative to the initial time when the
coordinates are specified by the barrier ensemble.  
If the initial conditions $\Delta
x_\text{u}(0)$ and $\Delta x_\text{s}(0)$ in the reactive degree of freedom
are prescribed, the reaction time can be calculated explicitly. The
dynamics of the reactive degree of freedom is given by
\begin{equation}
  \label{relDyn}
  \Delta x_\text{u}(t) = \Delta x_\text{u}(0) e^{\epsilon_\text{u}t} \;,\qquad
  \Delta x_\text{s}(t) = \Delta x_\text{s}(0) e^{\epsilon_\text{s}t} \;.
\end{equation}
The dividing surface is characterized by the condition $\Delta x=0$, which
can be rewritten in relative coordinates as $\Delta x_\text{u}=-\Delta
x_\text{s}$. The reaction time $\Delta t^\ddag$ at which this condition is
satisfied is easily found to be
\begin{align}
  \label{reactTime}
  \Delta t^\ddag&= \frac{1}{\epsilon_\text{u}-\epsilon_\text{s}}\,
     \ln\frac{-\Delta x_\text{s}(0)}{\Delta x_\text{u}(0)}\nonumber\\
   &= \frac{1}{\epsilon_\text{u}-\epsilon_\text{s}}\,
      \ln\frac{\Delta v_x(0)-\epsilon_\text{u} \Delta x(0)}
              {\Delta v_x(0)-\epsilon_\text{s} \Delta x(0)} \;.
\end{align}
It is defined for all initial conditions that are either forward- or
backward-reactive. For a forward-reactive trajectory, $\Delta
v_x(0)>0$. Because $\epsilon_\text{u}>0$ and $\epsilon_\text{s}<0$, it can
easily be seen from Eq.~(\ref{reactTime}) that $\Delta t^\ddag>0$ if $\Delta
x(0)<0$, as it should be for trajectories that start on the reactant side
of the dividing surface and are still to cross it. Similarly, a trajectory
with $\Delta x(0)>0$ is already on the product side, and its reaction time
is negative.  A backward-reactive trajectory, on the other hand, has an
initial velocity $\Delta v_x(0)<0$. In this case, $\Delta t^\ddag<0$ if
$\Delta x(0)<0$ and $\Delta t^\ddag>0$ if $\Delta x(0)>0$.

If the initial position $\Delta x(0)$ is fixed, the reaction
time~(\ref{reactTime}) tends to zero as $\Delta v_x(0)\to\infty$:
Trajectories with large initial velocities cross the barrier fast. On the
other hand, as the separatrices that bound the reactive region are
approached, i.e. $\Delta v_x(0)\to\epsilon_\text{u}\Delta x(0)$ if $\Delta
x(0)>0$ or $\Delta v_x(0)\to\epsilon_\text{s}\Delta x(0)$ if $\Delta
x(0)<0$, trajectories keep barely enough energy to cross the barrier, and
their reaction times tend to $+\infty$ or $-\infty$, respectively.

Once the reaction time is given as a function of initial conditions, 
the distribution for the forward- or backward-reactive
part of the barrier ensemble~(\ref{BarrDist}) is
readily obtained. 
In the former case, its probability distribution function is given by
\begin{multline}
  \label{tDist}
  p(\Delta t) = \frac{1}{P_\text{f}}
    \int d\Delta x \int_{\Delta v_x>\Delta v_{x,\text{min}}} d\Delta v_x
    \,\times\\
    \int d\Delta y \int d\Delta v_y\;
    f_\text{rel}(\Delta x,\Delta y,\Delta v_x,\Delta v_y)\,\times\\
    \delta(\Delta t-\Delta t^\ddag(\Delta x,\Delta v_x,\Delta y,\Delta v_y))\;.
\end{multline}
The normalization factor $1/P_\text{f}$ accounts for the fact that only the
forward-reactive part of the ensemble contributes to the distribution.

The distribution function~(\ref{tDist}) can in its most convenient form be
written in terms of the dimensionless scaled time
$\Delta\tau^\ddag=(\epsilon_{\text{u}}-\epsilon_{\text{s}}) \Delta
t^\ddag$. It then reads
\begin{align}
  \label{tauDist}
  \tilde p(\Delta\tau^\ddag)=&\frac{1}{\epsilon_\text{u}-\epsilon_\text{s}}\,
           p(\Delta\tau^\ddag/(\epsilon_\text{u}-\epsilon_\text{s}))
           \nonumber \\
         =&\frac{|r|}{\sqrt{\pi}P_{\text{f}}}\,
          \frac{e^{\Delta\tau^\ddag}}{(1-e^{\Delta\tau^\ddag})^2}
          \,\times\nonumber\\
         &\exp\left\{ -\left(\frac{r}{1-e^{-\Delta\tau^\ddag}}+w\right)^2
              \right\} \;,
\end{align}
where
\begin{equation}
  r=\frac{q_\alpha^\ddag(0)\,
          (\epsilon_{\text{u}}-\epsilon_{\text{s}})}
         {\sqrt{2\,k_{\text{B}}T}} \;,
\end{equation}
\begin{equation}
  w=\frac{v_\alpha^\ddag(0)-\epsilon_\text{u} q_\alpha^\ddag(0)}
         {\sqrt{2\,k_\text{B}T}}\;.
\end{equation}
The reaction probability $P_\text{f}$ can be written in terms of $r$ and
$w$ as
\begin{equation}
  P_\text{f} = \begin{cases}
                  \frac 12\operatorname{erfc}(r+w) & 
                     :\quad r>0 \;,\\
                  \frac 12\operatorname{erfc}(w)  & 
                     :\quad r<0 \;.
                 \end{cases}
\label{eq:normr}
\end{equation}
The valid range of $\Delta\tau^\ddag$ is $0<\Delta\tau^\ddag<\infty$ if
$q_\alpha^\ddag(0)>0$ and $-\infty<\Delta\tau^\ddag<0$ if
$q_\alpha^\ddag(0)<0$. The distribution function~(\ref{tauDist}) is
normalized so that its integral over that range is one.  Remarkably, the
distribution depends only on the two parameters $r$ and $w$, even though
the system dynamics and the distribution of initial conditions are
determined by the five parameters $\omega_{\text{b}}$, $\gamma$,
$T$, $q_\alpha^\ddag(0)$ and $v_\alpha^\ddag(0)$.

In a similar manner, the distribution of reaction times can be computed for
the backward-reactive part of the ensemble. The result is again given
by Eq.~(\ref{tauDist}), except that the valid range is now
$-\infty<\Delta\tau^\ddag<0$ if $q_\alpha^\ddag(0)>0$ and
$0<\Delta\tau^\ddag<\infty$ if $q_\alpha^\ddag(0)<0$. To obtain the proper
normalization, the reaction probability $P_\text{f}$ in the prefactor
of Eq.~(\ref{tauDist}) must be replaced by the backward-reaction probability
$P_\text{b}$, which in terms of the scaled parameters reads
\begin{equation}
  P_\text{b} = \begin{cases}
                 \frac 12\operatorname{erfc}(-w) & 
                     :\quad r>0 \;,\\
                  \frac 12\operatorname{erfc}(-r-w)  & 
                     :\quad r<0 \;.
               \end{cases}
\label{eq:normb}
\end{equation}

\begin{figure}
 \centerline{\includegraphics[width=\columnwidth]{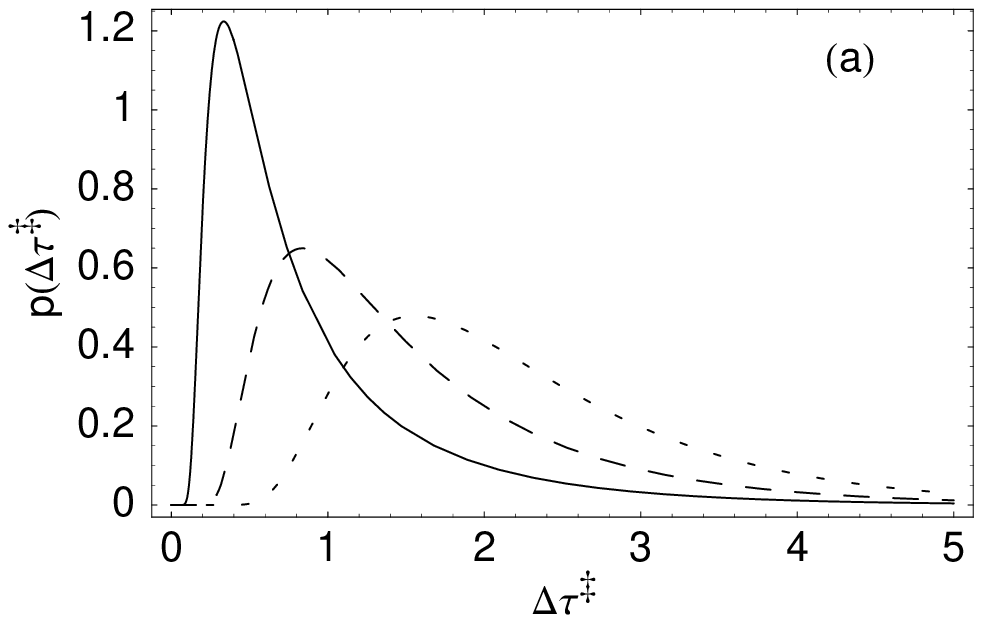}}
 \centerline{\includegraphics[width=\columnwidth]{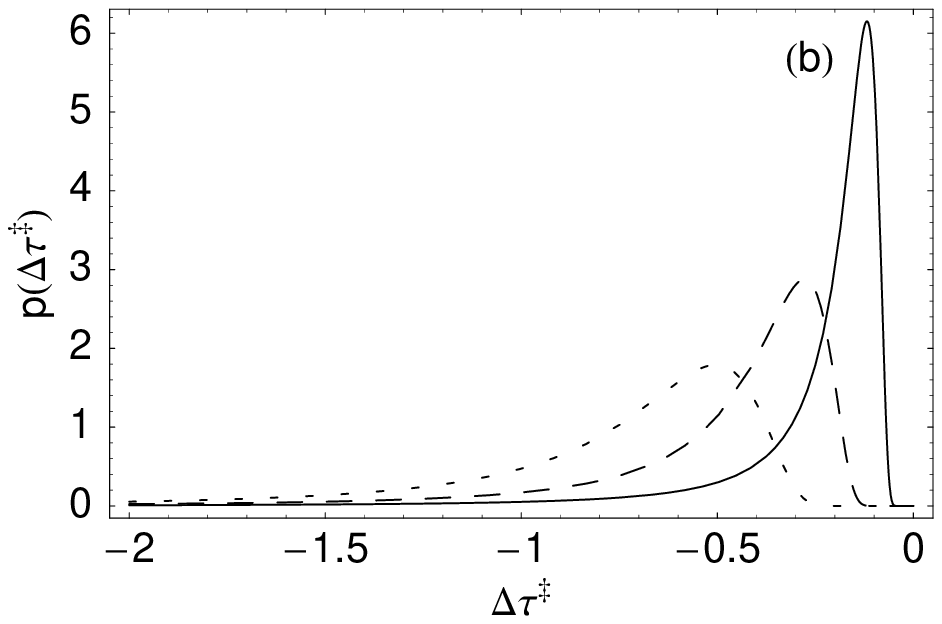}}
\caption{\label{fig:dist}
  The distribution~(\ref{tauDist}) of reaction times (a) for $w=1$ and $r=0.2$
  (solid), $r=0.5$ (dashed) and $r=1$ (dotted), (b) for $w=-1$ and $r=-0.2$
  (solid), $r=-0.5$ (dashed) and $r=-1$ (dotted).
}
\end{figure}

As can be seen from Fig.~\ref{fig:dist}, the reaction-time
distribution~(\ref{tauDist}) is highly asymmetric around its peak. The
probability distribution function is flat at $\tau=0$, where all of its
derivatives are zero. For large $|\tau|$, it decays exponentially like
\begin{equation}
  \label{pAsympt}
  \tilde p(\Delta\tau^\ddag)\approx
     \begin{cases}
         \frac{r\,e^{-(r+w)^2}}{\sqrt{\pi}\,P_\text{f}}\,e^{-\Delta\tau^\ddag} \\
         \hspace{4em}\text{if }
             q_\alpha^\ddag(0)>0, \quad \Delta\tau^\ddag\to+\infty \;, \\[2ex]
         \frac{r\,e^{-w^2}}{\sqrt{\pi}\,P_\text{f}}\,e^{\Delta\tau^\ddag} \\
         \hspace{4em}\text{if }
          q_\alpha^\ddag(0)<0, \quad \Delta\tau^\ddag\to-\infty\;.
     \end{cases}
\end{equation}
Because the distribution is so highly asymmetric, the average reaction time
will be significantly larger than the most probable reaction time that is
indicated by the maximum of the distribution function.

\section{Numerical results}\label{sec:results}

As soon as the anharmonicities of the potential in a realistic chemical
system have to be taken into account, the equations of
motion can no longer be solved analytically, and recourse must be taken to
numerical methods.
In what follows, the initial conditions, at $t=0$, are chosen
from the distribution in Eq.~(\ref{BarrDist}).
All trajectories are evolved forward and backward in time to $t=\pm
T_\text{int}/2$ using the stochastic integration algorithm introduced by
Ermak and Buckholz \cite{ermak80,allen87}. For the backward propagation,
the integration scheme was modified as described in the appendix.
In a conventional calculation of the exact rate expresssion,
reactive trajectories are identified according to the positions 
they attain at the start and end of the integration interval: 
Trajectories that at $t=-T_\text{int}/2$ and
$t=+T_\text{int}/2$ are located on opposite sides of the space-fixed
dividing surface $x=0$ are classified as forward- or backward-reactive;
others are classified as nonreactive. 
This criterion, however, is only reliable if the
total integration time $T_\text{int}$ is sufficiently large. At short
times, recrossings of the dividing surface introduce unavoidable errors. 

An alternative criterion for the identification of
reactive trajectories is obtained
if the space-fixed dividing surface is replaced by 
the moving TS surface described above.
In the most naive implementation, trajectories can be classified as
reactive if they are on opposite sides of the moving TS surface at $t=\pm
T_\text{int}$. 
If the moving-TS-surface algorithm is used instead,
$T_\text{int}$ can be reduced by as much as
a factor of 2 while still obtaining nearly accurate results. 
In addition, given
that the moving TS surface is exactly free of recrossings in the harmonic
limit and approximately so in an anharmonic potential, the integration can
be stopped as soon as a trajectory crosses the moving surface: There is no
need to follow the trajectory further and check for recrossings. Therefore,
when the moving TS surface is used, the actual integration time will on
average be much smaller than the nominal integration time $T_\text{int}$.

The reliability of this identification is illustrated below using
the two-dimensional saddle point potential of Eq.~\ref{eq:pot}
with and without anharmonicity, $k$.
In all of the numerical calculations, the units are chosen for simplicity such
that $k_\text{B}T=1$.
The friction is isotropic, with $\gamma=0.2$ in these units, 
and selected so as to be near the turnover between the energy- and
space-diffusion limited regimes. 
Although most of the calculations assume the same fixed noise sequence,
averages of the forward and backward reaction probabilities over the 
noise are also shown below.
In the former, the number of trajectories is fixed at $N_{\rm t}=15\,000$,
which is large enough to make statistical errors negligible. 
In the single-noise calculations on the two-dimensional harmonic barrier,
the transverse frequency $\omega_y=1.5$, and 
the barrier frequency is set to $\omega_x=1.0$.
The latter is reduced to
$\omega_x=0.75$ for the noise-averaging and in the nonlinear 
cases in order to accentuate the nonlinear coupling.

\subsection{Harmonic Systems}
\label{ssec:resultsHarm}

\begin{figure}
 \begin{center}
 \includegraphics*[width=\columnwidth]{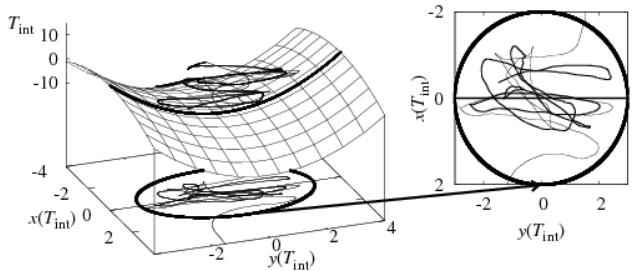}
 \end{center}
\caption{
 The evolution of a member of the ensemble 
 and the Transition State Trajectory depicted as the gray and black lines,
 respectively.
 The underlying potential is included, and the 
 fixed transition state $x=0$ is highlighted by the heavy black line.
 The time-independent projection is shown
 on the base of the figure.
 The sample trajectory is backward-reactive since it is a reactant in the
 future and product in the past.
 As can be seen, the Transition State Trajectory 
 remains in the vicinity of the  barrier for all times.
}
\label{fig:traj}
\end{figure}

\begin{figure}
 \begin{center}
 \includegraphics*[width=\columnwidth]{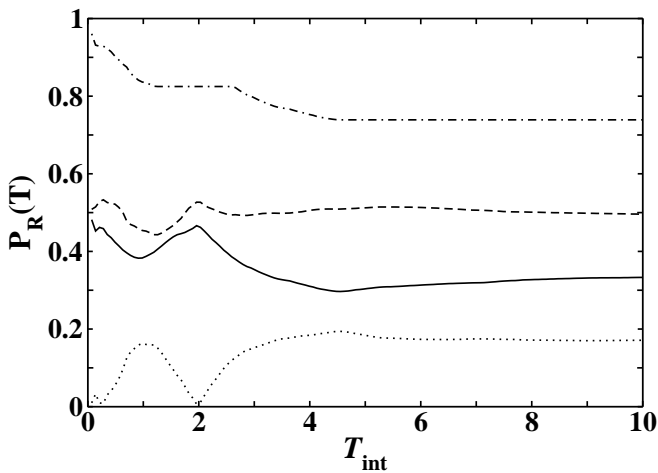}
 \end{center}
\caption{
Reaction probabilities calculated using the fixed dividing surface displayed
as a function of total integration time.
The fractions of forward-reactive,
backward-reactive, and nonreactive 
trajectories are shown as the solid, dashed, and dotted lines,
respectively.
The dash-dotted line represents the fraction of trajectories
that cross the surface only once at $T_\text{int}=0$.
In these simulations, $N_{\rm t}=15000$ trajectories were integrated,
the friction constant $\gamma=0.2$, and the barrier frequency is $\omega_x=1$. 
}
\label{fig:react}
\end{figure}

A typical reactive trajectory and the TS trajectory in the 
harmonic limit ($k=0$), are shown in Fig.~\ref{fig:traj}. 
Clearly, the space-fixed dividing surface $x=0$, 
in contrast to the moving TS surface, is crossed many times.
The respective percentages of trajectories 
classified either as reactive and nonreactive 
using the fixed dividing surface
are displayed as a function of integration time in Fig.~\ref{fig:react}.  
Because all trajectories start on the dividing surface, 
at very short times, every trajectory is classified as either forward- or
backward-reactive. 
Subsequent recrossings of the transition state result in
transient fluctuations of the reaction probabilities that slowly approach
the true, long-time values.  Figure~\ref{fig:react} also shows the
percentage of trajectories that are nonreactive as well as those that cross
the fixed dividing surface only once at $T_{\rm int}=0$.  
The latter comprise the majority of the reaction events, whereas the
percentage contributed by reactive trajectories is comparatively small. 
Nevertheless, the fluctuations in the computed
reaction probabilities that are caused by recrossings are considerable.

\begin{figure}
 \begin{center}
 \includegraphics*[width=\columnwidth]{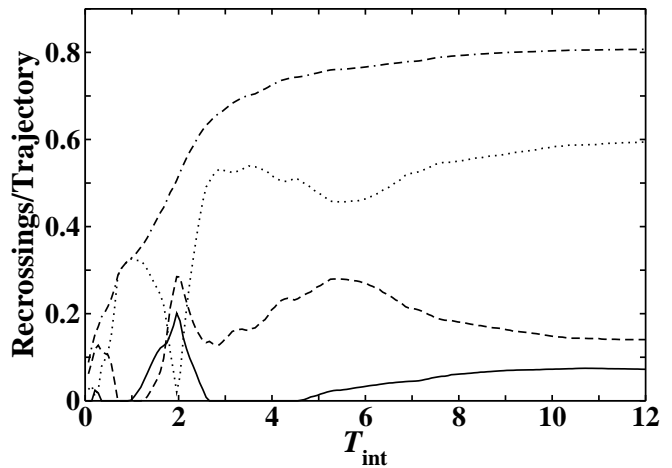}
 \end{center}
\caption{
The number of {\em recrossings} (dot-dashed line) of the fixed 
transition state normalized by the total number of trajectories.
The components of the total that resulted in a 
forward-reactive or backward-reactive trajectory are
shown as the solid and dashed lines, respectively.
Recrossings that resulted in a nonreactive trajectory 
are displayed as the dotted line.
The simulation parameters are the same as in Fig.~\ref{fig:react}.
}
\label{fig:crossings}
\end{figure}

Because recrossings are crucial to the performance of the algorithm, 
it is instructive to analyze them in more detail.
Figure~\ref{fig:crossings} shows the average number of recrossings per
trajectory as a function of the total integration time. The trivial 
crossing of the dividing surface that all trajectories undergo at $t=0$ is
not included.  The number increases monotonically as the trajectories cross
and recross the transition state. Eventually, it reaches a plateau as they
leave the barrier region and are lost into either the product or reactant
states.
In addition, Fig.~\ref{fig:crossings} decomposes the total number of
recrossings into those recrossings that occur on trajectories that are
found to be forward-reactive, backward-reactive or nonreactive at the given
integration time. Because the classification
of a particular trajectory can change with increasing integration time,
these contributions are not monotonic. Most prominently, as the number of
nonreactive trajectories decreases almost to zero at $T_\text{int}\approx
2$ (see Fig.~\ref{fig:react}), the contribution of nonreactive
trajectories shows the same behavior.
For large integration times, the largest contribution to recrossings stems
from nonreactive trajectories, which are bound to recross the dividing
surface at least once. In fact, a comparison of Fig.~\ref{fig:react} and
Fig.~\ref{fig:crossings} reveals that nonreactive trajectories on average
recross more than three times before they finally leave the barrier region.
Most of the reactive trajectories, by contrast, do not recross, and their
contribution to the recrossing statistics is much smaller. Asymptotically,
both forward and backward reactive trajectories recross on average
approximately 0.25~times.

\begin{figure}
 \begin{center}
 \includegraphics*[width=\columnwidth]{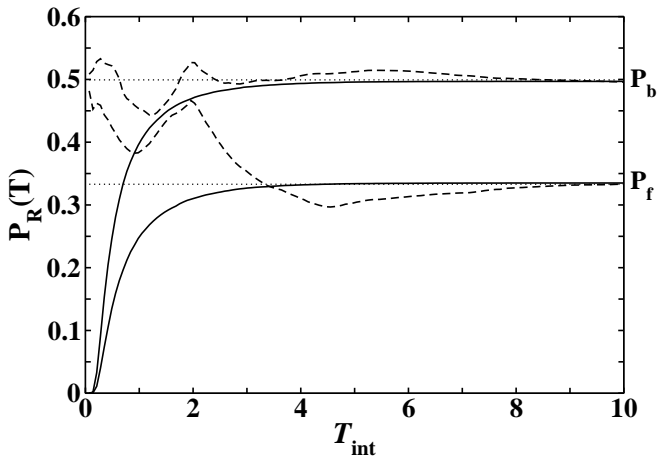}
 \end{center}
\caption{
Reaction probabilities as a function of integration time calculated using
the moving (solid curves) or the 
fixed (dashed curves) dividing surface.
The upper set of curves represents forward-reactive probabilities, 
with the lower set depicting the corresponding back reactions. 
The dotted curves indicate the asymptotic values $P_\text{f}=0.3332$ and
$P_\text{b}=0.4993$ calculated from Eq.~(\ref{eq:normr}) or~(\ref{eq:normb}),
respectively.
The simulation parameters are the same as in Fig.~\ref{fig:react}.
}
\label{fig:TST}
\end{figure}

The dynamics is greatly simplified if the moving TS surface is used
instead of the fixed one. Reaction probabilities computed using either
surface are compared in Fig.~\ref{fig:TST}. Because the trajectories
start at a distance from the moving TS surface, the corresponding rates are
zero for short integration times. They then steadily increase toward the
true long-time probabilities. Since the dividing surface cannot be
recrossed, the asymptotic values are approached monotonically.
The erratic fluctuations of the computed reaction probability that 
the fixed surface produces are absent if the moving TS surface is used, so
that a strict lower bound for the reaction probability is obtained even for
very short integration times. 
In quantitative terms,
the moving TS surface identifies a trajectory as reactive if its
reaction time $\Delta t^\ddag$ lies within the integration interval, so
that the finite-time reaction probability for a forward reaction is given
by
\begin{equation}
  \label{PrFinite}
  P_\text{f}(T_\text{int}) = P_\text{f} \times
      \operatorname{Prob}\left\{|\Delta t^\ddag| < \frac{T_\text{int}}2
                         \right\} \;,
\end{equation}
and a similar expression for the backward-reaction probability. The
reaction probabilities computed from the moving TS surface are therefore
determined by the distribution~(\ref{tauDist}) of reaction
times. The convergence toward the long-time probability
is described by the long-time tail~(\ref{pAsympt}) of the
reaction-time distribution and is exponentially fast. Indeed,
Fig.~\ref{fig:TST} shows that reaction probabilities computed using the
moving TS surface converge much faster than those obtained from the fixed
surface.
Moreover, in cases 
such as the current problem, in which the separatrices
between reactive and non-reactive trajectories are known exactly,
the reaction probabilities $P_\text{f}$ and $P_\text{b}$ can be computed 
\emph{a priori}, without having to perform any numerical simulations.  The
values obtained from Eqs.~(\ref{eq:normr}) and~(\ref{eq:normb}) are
also indicated in Fig.~\ref{fig:TST}. They agree precisely with the
asymptotic probabilities obtained from the simulation.
Thus, the moving TS surface can provide 
accelerated convergence in the rate for finite-time computations
for linear 
problems.  

\begin{figure}
\includegraphics*[width=\columnwidth]{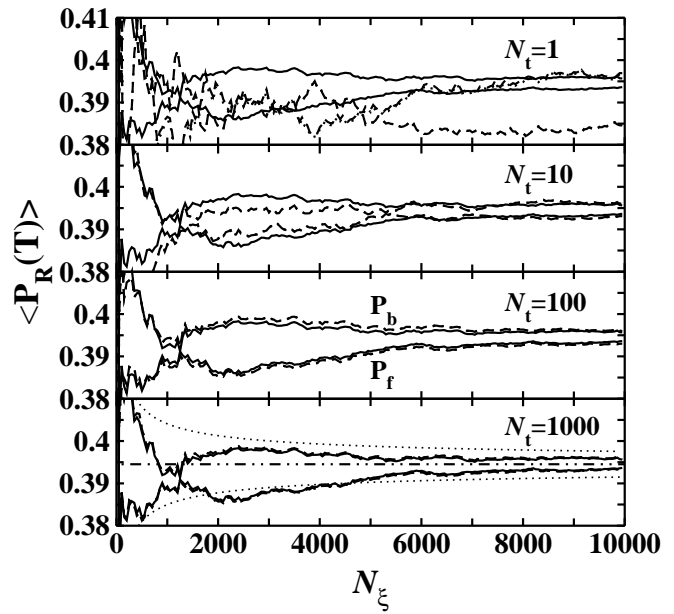}
\caption{
The reaction probabilities averaged over different instances of the noise
on the harmonic potential for four different values of the 
number of trajectories ($N_{\rm t}$) in the ensemble.
The solid lines depict the results predicted by 
Eqs.~(\ref{eq:normr}) or ~(\ref{eq:normb}).
The dashed and dotted lines are the results obtained using
the respective fixed or moving dividing surfaces.
In the harmonic case, these two surfaces provide the same results and
are indistinguishable.
For the case of $N_{\rm t}=1000$ the light dotted lines display the 
$95\%$ confidence interval with respect to the number of noise sequences 
sampled ($N_{\xi}$).
The simulation parameters are the same as those
defined in Fig.~\ref{fig:coupled}.
}
\label{fig:harm_avg}
\end{figure}

\begin{figure}
\includegraphics*[width=\columnwidth]{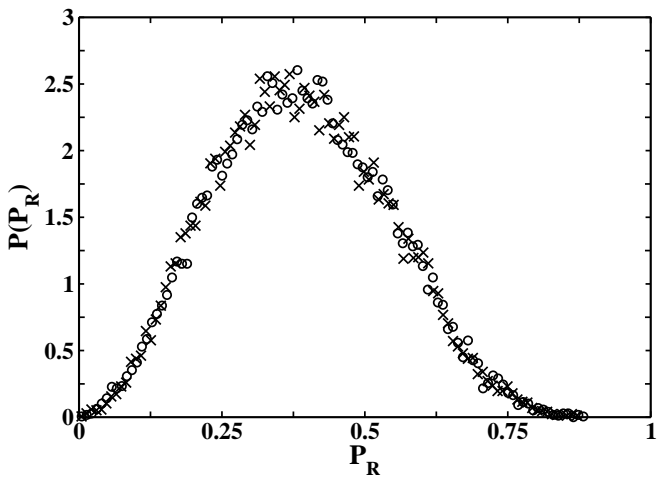}
\caption{
The distribution of reaction probabilities in the harmonic limit calculated  
from Eq.~(\ref{eq:normr}) or~(\ref{eq:normb}) from $N_{\xi}=20000$ 
different instances of the noise.
The x-symbols display the results for forward-reactive probabilities and
the o-symbols are for backward-reactive. 
The simulation parameters are the same as those
defined in Fig.~\ref{fig:coupled}.
}
\label{fig:dist_avg}
\end{figure}

The analytic reaction probabilities, Eqs.~(\ref{eq:normr}) and~(\ref{eq:normb}),
for the harmonic barrier represent the limiting values that are obtained
for one instance of the noise using a large number $N_{\rm t}$ of
trajectories.
To obtain a macroscopically observable reaction probablity, one has to
average these results over a large number $N_{\xi}$ of realizations of the noise.
That average cannot be obtained
analytically, but it can be easily calculated by a numerical quadrature.
It provides a useful benchmark for the
convergence of the computational schemes with respect to $N_{\rm t}$ and $N_{\xi}$. 
Figure~\ref{fig:harm_avg} illustrates the forward and backward reaction
probabilities, averaged over $N_{\xi}$ realizations of the noise, as a function
of $N_{\xi}$ and for different values of $N_{\rm t}$. 
The solid and dashed curves are obtained if reactive trajectories are
identified through the criteria provided by the fixed and the moving TS
surfaces, respectively.
As expected for a symmetric barrier, forward and backward reaction
probabilities converge toward the same limit. 
Moreover, the distributions
of forward and backward reaction probabilities agree, 
as shown in Fig.~\ref{fig:dist_avg}. 
For large $N_{\rm t}$, the results in Fig.~\ref{fig:harm_avg} 
agree with the analytic value displayed
as the dot-dashed curve in the figure's bottom panel.
Therein, dotted curves are used to indicate
the 95\% confidence interval to further illustrate
that the simulations are converging toward the correct limit as expected.

The simulation results in Fig.~\ref{fig:harm_avg} 
that employ the conventional criterion for identifying trajectories
have been computed using the large integration time
$T_\text{int}=21.5$, 
to illustrate the exact results within the error
bars of the number average.
However, it should be clear from Fig.~\ref{fig:react} that the
moving-TS-surface criterion often identifies reactive trajectories
in less than half this time, and once so identified a trajectory
does not need to be integrated further.
Given that the calculation of the moving surface itself 
---which amounts to the calculation of the TS trajectory---
takes roughly as much computational
effort as the integration of an ensemble trajectory, 
computational savings can thus be obtained from the use of 
the moving surface whenever the number $N_{\rm t}$
of trajectories per noise sequence is larger than 2.

\subsection{Nonlinear Systems}
\label{ssec:resultsNonlin}

The true test for the usefulness of the moving transition state lies
in its ability to
identify reactive trajectories beyond the linear regime.  If nonlinearities
are present, the relative coordinate~(\ref{relCoord}) does not achieve a
complete separation of the relative motion from the motion of the TS
Trajectory. Therefore, the moving dividing surface will not strictly be
free of recrossings. However, if the nonlinearities are weak, recrossings
can be expected to be rare. In these cases, the moving dividing surface
will be recrossing-free to a useful approximation. Indeed, our results
indicate that its advantages over a fixed dividing surface persist well
beyond the harmonic limit.

\begin{figure}
 \begin{center}
 \includegraphics*[width=\columnwidth]{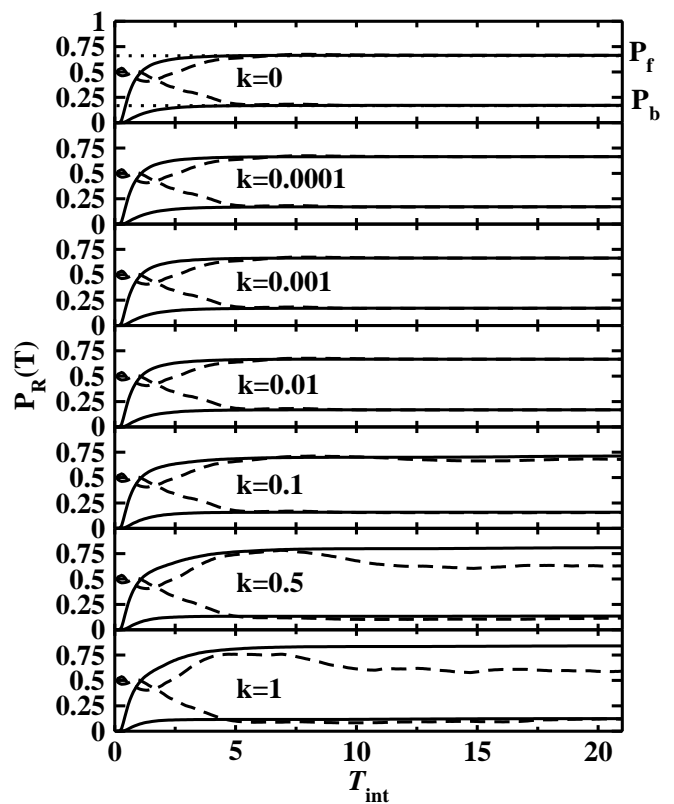}
 \end{center}
\caption{
Reaction probabilities as a function of integration time
calculated using the moving (solid line) or fixed dividing 
surface (dashed line) for
various values of the coupling constant.
The time step has been reduced to $8\times 10^{-6}$ for
convergence and
the barrier frequency changed to $\omega_x=0.75$
to accentuate the nonlinearity.
For the case $k=0$, the results for the reaction
probabilities as calculated from Eq.~(\ref{eq:normr}) or ~(\ref{eq:normb})
are included as the dotted lines. 
}
\label{fig:coupled}
\end{figure}

We investigate the performance of the moving dividing surface in the
example of the potential~(\ref{eq:pot}), with the coupling constant $k$ now
taking non-zero values.
The reaction probabilities for several different values of~$k$
are displayed in Fig.~\ref{fig:coupled}. To accentuate
the anharmonicity, the barrier frequency was reduced to $\omega_x=0.75$  to
allow
trajectories to spend more time in the barrier region before escaping.
For the transverse frequency, the value $\omega_y=1.5$ was retained.
Evidently, for sufficiently long integration times 
the moving transition state provides essentially the same
result as the fixed dividing surface for all values of the coupling
constant up to $k=0.1$.  
However, the reaction probabilities converge toward the long time
limit monotonically and much faster than those computed with the fixed dividing
surface. Therefore, the computational advantages that the moving surface
offers in the harmonic limit 
persist even in the presence of quite substantial nonlinearities.
Eventually, of course, the use of a moving
dividing surface based upon the harmonic approximation
ceases to be meaningful, as can be seen for $k=0.5$ and $k=1$.
For the specific instance of noise used in these calculations,
the results obtained from the moving and fixed dividing surfaces remain in 
agreement for the backward-reactive trajectories, whereas
a substantial difference arises for the forward-reactive trajectories. 
As is to be expected of any TST scheme, in these cases the moving dividing
surface overestimates the reaction probability because any trajectory
that crosses the surface is assumed to be reactive, whereas the possibility
of recrossings is neglected.
Although not shown, a different instance of the noise 
does not change the trends observed in Fig.~\ref{fig:coupled}.

\begin{figure}
 \begin{center}
 \includegraphics*[width=\columnwidth]{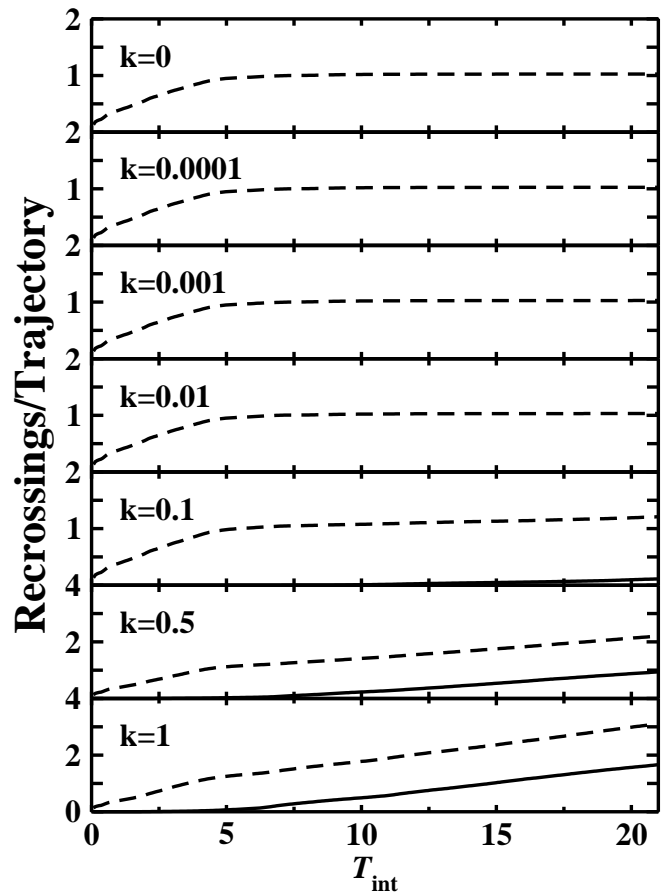}
 \end{center}
\caption{
The average number of recrossings of the moving transition state (solid lines)
and the fixed transition state (dashed line) normalized by
the total number of trajectories
for given values of the coupling constant.
The simulation parameters are the same as those
defined in Fig.~\ref{fig:coupled}.
The values for the moving transition state are too small to
be seen on the same scale in the top four panels.
\label{fig:recross}
}
\end{figure}

As in the harmonic limit, the computational advantages of the moving
dividing surface in systems with moderate nonlinearities stem from
the fact that it is approximately free of recrossings. 
This is illustrated in Fig.~\ref{fig:recross}.
The average number of recrossings per trajectory of the fixed transition
state exhibits similar behavior for small to moderate values of the
coupling constant. It
approaches approximately one recrossing per trajectory in the long-time limit.
In these cases, the number of recrossings of the moving dividing surface 
is so much smaller than the corresponding number for 
the fixed surface that it is not visible in the figure.
At larger coupling, the number of recrossings of the fixed dividing surface
does not converge to a finite long-time limit, but instead
increases linearly with the integration time.
The onset of a similar behavior occurs at 
approximately the same value of $T_{\rm int}$ for the moving dividing surface 
as well.

This increase in the number of recrossings for both the fixed and the
moving surfaces is caused by 
a small percentage of trajectories in the ensemble
that never leave the TS region for negative times, but rather get trapped in
an oscillation in the stable transverse degree of freedom $y$.
If the value of $y$ is sufficiently large, 
the reactive degree of freedom $x$ in the
potential~(\ref{eq:pot}) ceases to be unstable
but instead behaves as a harmonic oscillator with a (possibly large)
effective frequency $\tilde\omega_x^2=k\,y^2-\omega_x^2$.
As a result of these fast oscillations in the reactive degree of freedom, 
the dividing surfaces are crossed many times. 
This mechanism has been 
confirmed by a detailed trajectory analysis, which for brevity we do not show.
It is a rather peculiar feature of our model potential due to the 
presence of only one higher order coupling term in the 
potential~(\ref{eq:pot}). 
We would not expect such aberrant behavior
to arise in a typical system.

\begin{figure}
 \begin{center}
 \includegraphics*[width=\columnwidth]{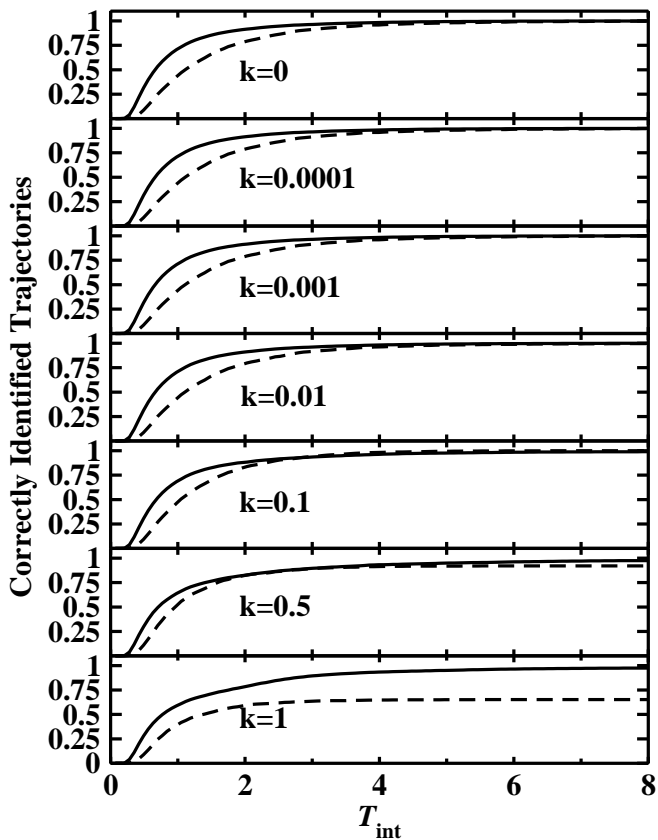}
 \end{center}
\caption{
The fraction of correctly identified trajectories according to
the moving transition state.
The correct identification of a trajectory is that defined by the
fixed transition state at 
the end of the simulation, $T_{\rm int}=21$.
The correctly identified backward-reactive or forward-reactive 
trajectories are displayed
as the solid and dashed lines, respectively.
The simulation parameters are the same as those
defined in Fig.~\ref{fig:coupled}.
}
\label{fig:percent}
\end{figure}

It is clear from Fig.~\ref{fig:coupled}
that for moderately strong anharmonicities the moving transition
state correctly identifies the overall number of reactive trajectories.
However, that number is a macroscopic observable, and
it is not immediately clear whether, on a microscopic level, 
individual reactive trajectories are identified correctly.
The fraction of trajectories that are identified correctly
by the moving transition state
is displayed in Fig.~\ref{fig:percent}, where the ``correct''
identification for a given trajectory has been assumed to be given by the
fixed dividing surface for a sufficiently long integration time.
In the cases of weak to moderate coupling, the classification obtained from
the moving dividing surface is correct for all trajectories, 
but, as expected, it begins to fail for coupling strengths around
$k=0.5$. 
The fact that the identification of the backward trajectories 
is poorer than that for the forward trajectories at large $k$ 
is not surprising.
The initial distribution ---particles located at the naive fixed 
transition state with forward velocity---  disfavors backward trajectories
which must recross the fixed TS at least twice more in order to 
reach the appropriate boundary conditions.
Nevertheless, Fig.~\ref{fig:percent} 
confirms that the favorable behavior of the
moving surface that is apparent in Fig.~\ref{fig:coupled} indeed reflects
a correct description of the underlying microscopic dynamics.

\begin{figure}
\includegraphics*[width=\columnwidth]{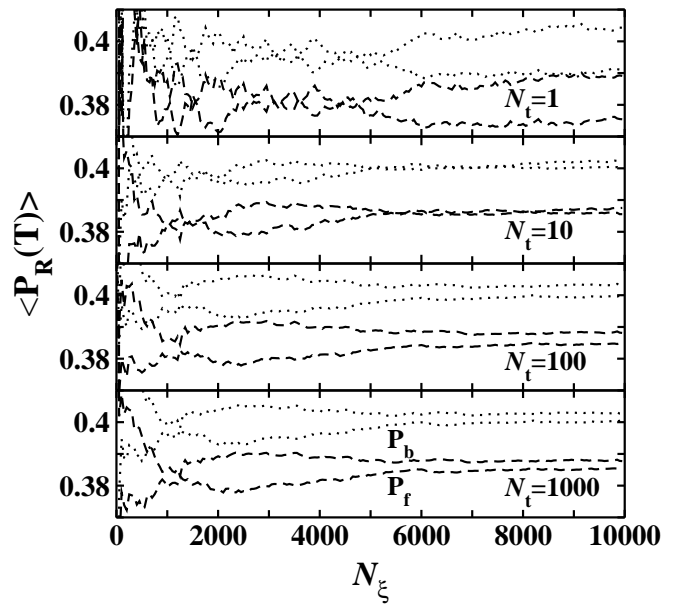}
\caption{
The reaction probabilities across an anharmonic potential 
(with a coupling of $k=0.1$) are shown as a smooth function 
of the number of different instances of the noise ($N_{\xi}$)
and a discrete function of the number of trajectories ($N_{\rm t}$) 
used to represent the ensemble average.
The dashed lines and dotted lines result from the use of
the fixed or moving dividing surfaces in the identification
of trajectories, respectively.
Note that the ordinate ranges over the very narrow interval between
.37 and .41, and hence the converged exact and approximate approaches 
are nearly equal.  
The simulation parameters are the same as those
defined in Fig.~\ref{fig:coupled}.
}
\label{fig:anh_avg}
\end{figure}

Figure~\ref{fig:anh_avg} displays the noise-averaged reaction probabilities
for the fixed and moving dividing surfaces for a coupling of $k=0.1$.  
The results display the same convergence behavior in both cases, except that
for the moving TS surface they are shifted to larger values by roughly $5\%$. 
This small error is due to the small percentage of trajectories that
recross the moving dividing surface, as seen for a particular instance of
the noise in Fig.~\ref{fig:recross}.  
Because the potential barrier
described by Eq.~\ref{eq:pot} is symmetric even for $k\ne 0$, the average
values of the forward and backward reaction probabilities~$P_{\rm f}$ and
$P_{\rm b}$ are equal.  
The simulation results converge rapidly, with
respect to both $N_{\rm t}$ and $N_{\xi}$, toward their limiting value.
These results demonstrate that the moving TS surface 
retains its reliability and its computational advantages for
moderate values of the anharmonicity upon noise averaging as well as for a
single instance of the noise.

\section{Concluding Remarks}

We have recently developed an analytic method for constructing
a time-dependent stochastic dividing surface that is
strictly free of recrossings \cite{dawn05,halcyon}.
In the present work, it has been shown that this 
moving dividing surface can be used to identify reactive
trajectories reliably in linear and nonlinear systems:
In the harmonic limit, the moving dividing surface attached to the TS
trajectory is strictly free of recrossings, while in more general 
(nonseparable) cases it is approximately so. 
The identification of reactive trajectories using the 
moving dividing surface has been seen in this 
article to be fairly accurate even in the presence of 
large anharmonic coupling.
It can be obtained
in roughly half the time that is required to confirm the nature of a
trajectory by numerically evolving it to its final state.

In several of the calculations presented in this article, 
observables have been calculated
for a particular instance of the noise while averaging
over the initial conditions of the subsystem.
In such restricted averages, the use of the moving surface reduces
the computational cost of the calculation by a factor of two or
more.
A typical average of an observable, however, requires one to include
multiple instances of the noise.  
When the average is performed using the machinery of the moving TS
surface, the TS trajectory must be generated for each instance of the noise.
If only one system trajectory is calculated for each noise sequence,
the computational effort to calculate both the sample trajectory
and the moving TS surface is about the same as calculating a single
(longer) trajectory.
Improved CPU performance can still be obtained if one
recognizes that the average should be taken by sampling several
trajectories for each noise sequence. 
Apart from the insight into the microscopic reaction dynamics 
that the moving dividing surface offers, it consequently
also provides computational
advantages in the calculation of macroscopic observables.
Moreover, the computation can be readily parallelized because
the algorithm is embarrasingly parallel when sampling across
the trajectories associated with a given noise sequence.
(Indeed, although not discussed explicitly in the text, 
the codes have been parallelized across several CPUs with near
linear scaling.)

In summary, we envision at least two approaches in which the
TS-trajectory criterion
for reactive trajectories will be useful in calculating reaction rates:
{\it (i)} In harmonic (or nearly harmonic) systems, the algorithm described
here provides a formally exact expression for the reaction probability
given a noise sequence.  
This term and related averages can be used to 
substantially reduce the required computational time 
because it 
limits the numerical effort to a sampling of the noise.
{\it (ii)} In arbitrary anharmonic systems, the criterion 
can be used to reduce the computational effort 
to calculate any correlation function 
---such as that in the reaction-rate expression---
that relies on the correct identification of reactive trajectories.
The rate expression and
other related observables that can take advantage
of the identification of reactive trajectories
will be calculated in future work.
As an illustration, the TS-trajectory criterion was seen in this work
to converge
the forward and backward reaction probabilities even
in a fairly anharmonic case.
Thus the central result of this work is:
the moving dividing surface can be
used reliably and efficiently to 
identify reactive trajectories.

\begin{acknowledgments}
This work was partly supported by the US National
Science Foundation and by the Alexander von Humboldt-Foundation.
The computational facilities at the CCMST have been supported under
NSF grant CHE 04-43564.
\end{acknowledgments}

\appendix

\section*{Appendix: The numerical integrator}

The simulations of the reaction dynamics presented in
Sec.~\ref{sec:results} require one to follow a stochastic trajectory
numerically from $t=0$ both forward in time to $t=T/2$ and backward in time
to $t=-T/2$. 
For the forward propagation, a standard stochastic
integration scheme \cite{ermak80,allen87} has been 
implemented. 
The backward integration requires special care if one wishes to follow the
same stochastic trajectory both forward and backward in time.
The modification of the integration scheme that is necessary to this end is
described here.

The forward numerical integrator for Langevin equations \cite{ermak80,allen87}
takes the form
\begin{eqnarray}
r(t+\delta t) & = & r(t) + c_1 v(t) + c_2 a(t) + \delta r \;,\\
v(t+\delta t) & = & c_3 v(t) + c_4 a(t) + c_5 a(t+\delta t) + \delta v
\;,
\end{eqnarray}
where $a(t)$ is the acceleration caused by the potential of mean force,
the $c_i$ are numerical coefficients that depend on
the time step $\delta t$ and the damping constant $\gamma$,
and the random variables $\delta r$ and $\delta v$ are sampled from a known
Gaussian distribution.
Time reversal in this algorithm can be obtained through a shift in time 
by $-\delta t$ so that $t$ becomes $t-\delta t$ 
and $t+\delta t$ becomes $t$. 
This replacement and a simple reorganization leads to
\begin{eqnarray}
\label{eq:pos}
r(t-\delta t) & = & r(t) - c_1 v(t-\delta t) - c_2 a(t-\delta t) - \delta r\;,\\
\label{eq:vel}
v(t-\delta t) & = & \frac{1}{c_3}\left[ v(t) 
                    - c_4 a(t-\delta t) - c_5 a(t) - \delta v \right]
\;.
\end{eqnarray}
The backward step~(\ref{eq:pos}) cannot be evaluated as it stands because
the acceleration $a(t-\delta t)$ depends on the
position $r(t-\delta t)$ that is yet to be determined.
To circumvent this problem, we substitute Eq.~(\ref{eq:vel})
into Eq.~(\ref{eq:pos}) to obtain
\begin{multline}
r(t-\delta t)=r(t)-\frac{c_1}{c_3}\left[ v(t) - c_5 a(t) -\delta v \right]\\
              -\delta r + \left(\frac{c_1c_4}{c_3}-c_2 \right)a(t-\delta t)
\;.
\label{eq:roft}
\end{multline}
When the acceleration $a(t-\delta t)$ is expressed in terms of the position
$r(t-\delta t)$ through the equation of motion, Eq.~(\ref{eq:roft}) becomes an
implicit equation for the positions $r(t-\delta t)$ at the earlier time.
For all but the simplest potentials, it cannot be solved explicitly.
Specifically, for the anharmonic potential~(\ref{eq:pot}),
\begin{equation*}
U(x,y)=-\frac{1}{2}\omega_x^2 x^2 +\frac{1}{2}\omega_y^2 y^2 + kx^2y^2 \;,
\end{equation*}
it leads to the coupled equation system
\begin{align}
x(t-\delta t) =&  X(t)+\left(\frac{c_1c_4}{c_3}-c_2\right) \times\nonumber\\
                  &\quad\left(\omega_x^2 x(t-\delta t) -
                         2kx(t-\delta t)y(t-\delta t)^2\right) \;,
                         \label{eq:collect1} \\ 
y(t-\delta t) =&  Y(t)-\left(\frac{c_1c_4}{c_3}-c_2\right) \times\nonumber\\
                   &\quad\left(\omega_y^2 y(t-\delta t) + 
                    2kx(t-\delta t)^2y(t-\delta t)\right)
\;, \label{eq:collect2}
\end{align}
where $X(t)$ and $Y(t)$ denote the contributions of the first three terms
in Eq.~(\ref{eq:roft}).
In the harmonic limit $k=0$, the two equations uncouple and can be solved
for simple explicit expressions for the position updates.
For nonzero $k$, Eqs.~(\ref{eq:collect1}) and~(\ref{eq:collect2})
represent an implicit integration scheme. 
It can be converted into an explicit method by rearranging the terms
into
\begin{eqnarray}
x(t-\delta t) & = & \frac{X(t)}{1-
                   \left(\frac{c_1c_4}{c_3}-c_2\right)
                   \left(\omega_x^2 -
                         2ky(t-\delta t)^2\right)} \;, \label{eq:expand1}\\
y(t-\delta t) & = & \frac{Y(t)}{1+
                    \left(\frac{c_1c_4}{c_3}-c_2\right)
                    \left(\omega_y^2 + 
                    2kx(t-\delta t)^2\right)}
\;.\label{eq:expand2}
\end{eqnarray}
The denominators in Eqs.~(\ref{eq:expand1}) and~(\ref{eq:expand2}) 
are updated using Eqs.~(\ref{eq:collect1}) and~(\ref{eq:collect2}), but 
the unknown corrections involving $a(t-\delta t)$ are neglected
because the
coefficient $c_1c_4/c_3-c_2$ is of second order in the time step $\delta
t$. This leads to
\begin{eqnarray}
x(t-\delta t)  & \approx & \frac{X(t)}{1-
                           \left(\frac{c_1c_4}{c_3}-c_2\right)
                           \left(\omega_x^2 - 2kY(t)^2 \right)} \;, \\
y(t-\delta t) & \approx & \frac{Y(t)}{1+
                          \left(\frac{c_1c_4}{c_3}-c_2\right)
                          \left(\omega_y^2 + 
                          2kX(t)^2\right)}
\label{eq:xoft}\;.
\end{eqnarray}
Finally, we insert these approximations into the right-hand sides of
Eq.~(\ref{eq:expand1}) and~(\ref{eq:expand2}) to obtain an explicit
integration scheme backwards in time.


\end{document}